\documentclass{aa}  

\newcommand{\mstar}{$M_{\star}\,$}
\newcommand{\lamre}{$\lambda_{R_e}$}
\newcommand{\frl}{$F_{\rm radio}\,$}
\defcitealias{Best12}{BH12}
\defcitealias{Graham19}{G19}
\defcitealias{Mulcahey22}{M22}
\newcommand{\radL}{$L_{150\rm\,MHz}\,$}
\usepackage{graphicx}
\usepackage{txfonts}
\usepackage{hyperref}

\begin{document}

   \title{MaNGA integral-field stellar kinematics of LoTSS radio galaxies: Luminous radio galaxies tend to be slow rotators}

   \author{Xuechen~Zheng
          \inst{1}
          \and
          Huub~R\"{o}ttgering
          \inst{1}
          \and 
          Arjen van der Wel
          \inst{2,3}
          \and
          Michele Cappellari
          \inst{4}
          }

   \institute{Leiden Observatory, Leiden University, PO Box 9513, NL-2300 RA Leiden, the Netherlands\\
              \email{zheng@strw.leidenuniv.nl}
              \and
              Sterrenkundig Observatorium, Department of Physics and Astronomy, Ghent University, Belgium
              \and
              Max-Planck Institut f\"{u}r Astronomie, K\"{o}nigstuhl 17, D-69117 Heidelberg, Germany
              \and
              Sub-department of Astrophysics, Department of Physics, University of Oxford, Denys Wilkinson Building, Keble Road, Oxford OX1 3RH, UK
             }

   \date{}

 
  \abstract
   {The radio jets of an active galactic nucleus (AGN) can heat up the gas around a host galaxy and quench star formation activity.
   The presence of a radio jet could be related to the evolutionary path of the host galaxy and may be imprinted in the morphology and kinematics of the galaxy.
   In this work, we use data from the Sloan Digital Sky  Survey’s Mapping Nearby Galaxies at Apache Point Observatory survey and the Low Frequency Array (LOFAR) Two-Metre Sky Survey as well as the National Radio Astronomy Observatory (NRAO) the {\it Karl G. Jansky} Very Large Array (VLA) Sky Survey and the Faint Images of the Radio Sky at Twenty Centimeter survey.
   We combine these integral field spectroscopic data and radio data to study the link between stellar kinematics and radio AGNs.
   We find that the luminosity-weighted stellar angular momentum \lamre is tightly related to the range of radio luminosity and the fraction of radio AGNs \frl present in galaxies, as high-luminosity radio AGNs are only in galaxies with a small \lamre, and the \frl at a fixed stellar mass decreases with \lamre.
   These results indicate that galaxies with stronger random stellar motions with respect to the ordered motions might be better breeding grounds for powerful radio AGNs.
   This would also imply that the  merger events of galaxies are important in the triggering of powerful radio jets in our sample.
   }
   %

   \titlerunning{MaNGA stellar kinematics of LoTSS radio galaxies}
   \authorrunning{X.~Zheng et al.}
   \maketitle\ 
   
%

\section{Introduction}\label{sec:intro}
A supermassive black hole (SMBH) actively accreting materials in a galaxy's centre is an active galactic nucleus (AGN). 
An AGN emits a large amount of energy and severely impacts the star-formation in its host galaxy \citep[see][ and reference therein]{Cattaneo09,Fabian12}.
Some AGNs are bright in radio bands and have prominent jet features. 
These AGNs are called `radio AGNs'. 
They usually reside in the centre of massive early-type galaxies \citep[ETG;][]{Condon78,Balick82,Best05b,Brown11}.
The radio jets can heat up the intergalactic materials around and therefore lower the supply of cooling gas for star-formation in their host galaxies \citep{Best06,Dunn06,McNamara07,Fabian12}.
Although radio AGNs can be important for the evolution of galaxies, it is unclear why some galaxies contain radio AGNs and others do not. 
To address this radio AGN-triggering problem, we investigate links between radio AGNs and the properties of their host galaxies. 

An intriguing fact about radio AGNs and their host galaxies is that the prevalence of radio AGNs is closely related to the morphology of galaxies.
\citet{Barisic19} show the fraction of radio AGNs in early-type galaxies increases with the optical axis ratio of galaxies using data from the Sloan Digital Sky Survey \citep[SDSS;][]{York00,Stoughton02}, the National Radio Astronomy Observatory (NRAO) the {\it Karl G. Jansky} Very Large Array (VLA) Sky Survey \citep[NVSS;][]{Condon02}, and the Faint Images of the Radio Sky at Twenty Centimeter survey \citep[FIRST;][]{Becker95}. 
This trend has also been confirmed for the high-luminosity radio AGNs ($L_{150\,\rm MHz}\gtrsim10^{23} \rm\,W\,Hz^{-1}$) in \citet{Zheng20} using the first data release of the LOFAR Two-metre Sky Survey \citep[LoTSS DR1;][]{Shimwell19}.
When taking the general changes in morphology as a function of stellar mass (\mstar) into account in these works, the radio AGN-morphology relation cannot be explained by the trend that more massive galaxies are more likely to be round \citep{vdWel09,Chang13}.
The radio AGN-morphology relation probably implies that high-power radio jets preferentially exist in galaxies with a merger-dominant history, which are more likely to be round.
This interpretation is in line with the results from the decomposition of the optical light profile of radio galaxies \citep{Wang16,Wang19} and deep imaging \citep{Tadhunter16,Pierce19}, where the host galaxies of high-luminosity radio AGNs show more post-merger features than their low-luminosity counterparts. 

We note that the link between radio AGNs and galaxy morphology is luminosity dependent.
\citet{Zheng20} indicated that the fraction of galaxies hosting low-luminosity ($L_{150\,\rm MHz}\lesssim10^{23} \rm\,W\,Hz^{-1}$) radio AGNs 
does not depend on the axis ratio of galaxies.
A large fraction of low-luminosity radio AGNs show no morphological signs of merging events and have disc-like components in optical images \citep[e.g.][]{Wang19,Pierce19}.
Therefore, it has been suggested that high- and low-luminosity radio AGNs may be triggered in different ways. 

The radio luminosity-morphology relation may be explained by the spin paradigm \citep{Blandford77,Wilson95,Sikora07,Fanidakis11,Chen21}.
As described in \citet{Blandford77}, the spin of an SMBH is an essential ingredient in the jet-launching process.
Numerical simulations have shown that both accretion processes and black hole mergers can change the spin of SMBHs \citep{Fanidakis11,Dubois14,Bustamante19}.
The black hole merger was believed to be an efficient way to make a high-spin SMBH \citep{Wilson95,Sikora07,Fanidakis11} and
led to the conclusion that merged SMBHs are more likely to launch radio jets.
Although this hypothesis has been challenged by both observational and numerical works \citep[][see Section \ref{sec:con} for more discussion]{Garofalo10,Reynolds13,Dubois14,Bustamante19,Sayeb21}, it is still reasonable to suggest that SMBHs in galaxies without mergers in their evolution path would have different spins than those in galaxies with a merger-driven history because the merging of galaxies would be followed by the merging of SMBHs.
%

When deep images are not available, it is difficult to verify whether a galaxy has a merger-dominant history with traditional imaging because of projection effects.
Integral field spectroscopic (IFS) observation, which can resolve the stellar velocity field in galaxies, presents a good way of overcoming this difficulty.
The pioneering work of \citet{Smith90} has shown that the stellar dynamics of galaxies can indeed provide valuable clues for studying the triggering of powerful radio AGNs.
Recently, IFS surveys such as the ATLAS$^{\rm 3D}$ survey \citep{Cappellari11}, the Calar Alto Legacy Integral Field Area survey \citep{Sanchez12}, and the Mapping Nearby Galaxies at APO \citep[MaNGA;][]{Bundy15} survey have greatly improved our understanding about the structure and stellar kinematics of galaxies.
These surveys have shown that ETGs can be separated into two categories, fast rotators (FRs) and slow rotators (SRs), based on stellar kinematics \citep[see ][and reference therein]{Cappellari16}.
The FRs have disc-like structures and rotation-dominated kinematics, while the SRs have no disc and random motion dominated kinematics.
The structure and stellar kinematics of these galaxies indicate their evolutionary paths.
Because disc structure can be easily preserved or rebuilt in gas-rich processes, such as gas accretion and wet mergers, SRs are most likely the products of major dry mergers, after which the stellar discs are disrupted and not enough gas is available to rebuild them. \citep{Bender92,Emsellem11,Bois11,Lauer12,Cappellari16}.
%

The FR-SR dichotomy seems to directly relate to the morphological difference in high- and low-luminosity radio AGNs.
In this work, we combine the observational results from MaNGA and LoTSS to investigate the stellar kinematics of high- and low-luminosity radio AGNs.
This can help verify the link between radio jets and a galaxy's evolution path.
This article is organised as follows:
We describe our sample in Section \ref{sec:data}.
The analyses and results are shown in Section \ref{sec:analysis}.
In Section \ref{sec:con}, we summarise the main findings of this work.
The cosmology adopted throughout the work is $H_0=70\rm\,km\,s^{-1}\,Mpc^{-1}$, $\Omega_{\rm M}=0.3$, $\Omega_{\Gamma}=0.7$.

\section{Data and sample}\label{sec:data}
The IFS data in this work are from the MaNGA Data Release 16 \citep[DR16;][]{Ahumada20}. 
%
The MaNGA survey uses the intergral field units (IFUs) equipped by the Baryon Oscillation Spectroscopic Survey (BOSS) spectrographs on the 2.5-metre Sloan Telescope \citep{Smee13}.
These IFUs have 19 to 127 hexagonally bundled 2" fibres corresponding to diameters of 12" to 32".
The final spectra provided by the MaNGA covers wavelengths from 360 to 1000\,nm, with a resolution of $R\sim$2000 \citep{Law16}.
The MaNGA DR16 contains a total of 4597 galaxies from an extended version of the NASA-Sloan Atlas (NSA\footnote{\url{http://www.nsatlas.org}}). 
These galaxies are selected to have a flat number density distribution in the {\it i}-band absolute magnitude within $0.01<z<0.15$ \citep{Wake17}. 
This selection strategy ensures that the galaxy sample has a flat stellar mass distribution between $10^9-10^{12}\,M_\odot$.
%
Galaxies in the primary and secondary samples of MaNGA are mapped by IFUs out to 1.5 and 2.5 effective radii ($R_e$), respectively.
%

The kinematic and morphological parameters of the MaNGA galaxies were measured by \citet[]{Graham19} (hereafter `G19') and tabulated in \citet{Bevacqua22}.
These galaxies were flagged based on the criteria in \citet{Graham18} to exclude problematic data, merging sources that have hard to define morphological parameters and sources too small with respect to the point spreading function of the Sloan Telescope.
%
After the exclusion of these sources, the rest of the galaxies formed the `clean' sample, as defined in \citetalias{Graham19}, which comprises 4003 sources.
These galaxies all have reliable kinematic and morphological measurements.
The galaxies used in our analysis are all from the clean sample.

To study the stellar kinematics of the galaxies, \citetalias{Graham19} measured the luminosity-weighted stellar angular momentum  within one $R_e$ ($\lambda_{R_e}$), defined in \citet{Emsellem07},
\begin{equation}
  \lambda_{R_e} = \frac{\sum_{n=1}^N F_nR_n|V_n|}{\sum_{n=1}^{N}F_nR_n\sqrt{V_n^2+\sigma_n^2}},  
\end{equation}
where $F_n$, $R_n$, $V_n$, and $\sigma_n$ are respectively the flux, projected radius, projected stellar velocity, and velocity dispersion of the $n$-th spatial bin.
The typical error of \lamre in the catalogue is less than 0.05 \citep{Graham18}. 
The galaxies with ellipticity $\epsilon<0.4$ and \lamre$<0.08+\epsilon/4$ were classified as SRs, while the others were classified as FRs \citep{Emsellem11,Cappellari16,Graham18}.

To identify the radio AGNs in the MaNGA sample, we used the data from the LoTSS Data release two \citep[DR2;][]{Shimwell19}. 
The LoTSS DR2 covers 27\% of the northern sky at 120--169MHz.
This survey achieves a median sensitivity of 83$\,\mu\rm Jy\,beam^{-1}$ and an astrometric accuracy of 0.2" at 6" resolution for bright sources.
A total of 4396228 radio sources were detected in LoTSS DR2.
Because of the high sensitivity and the large sky coverage, LoTSS DR2 is expected to have many low-luminosity radio AGNs, which makes it a good dataset for our work.

\begin{figure}
    \includegraphics[width=\linewidth]{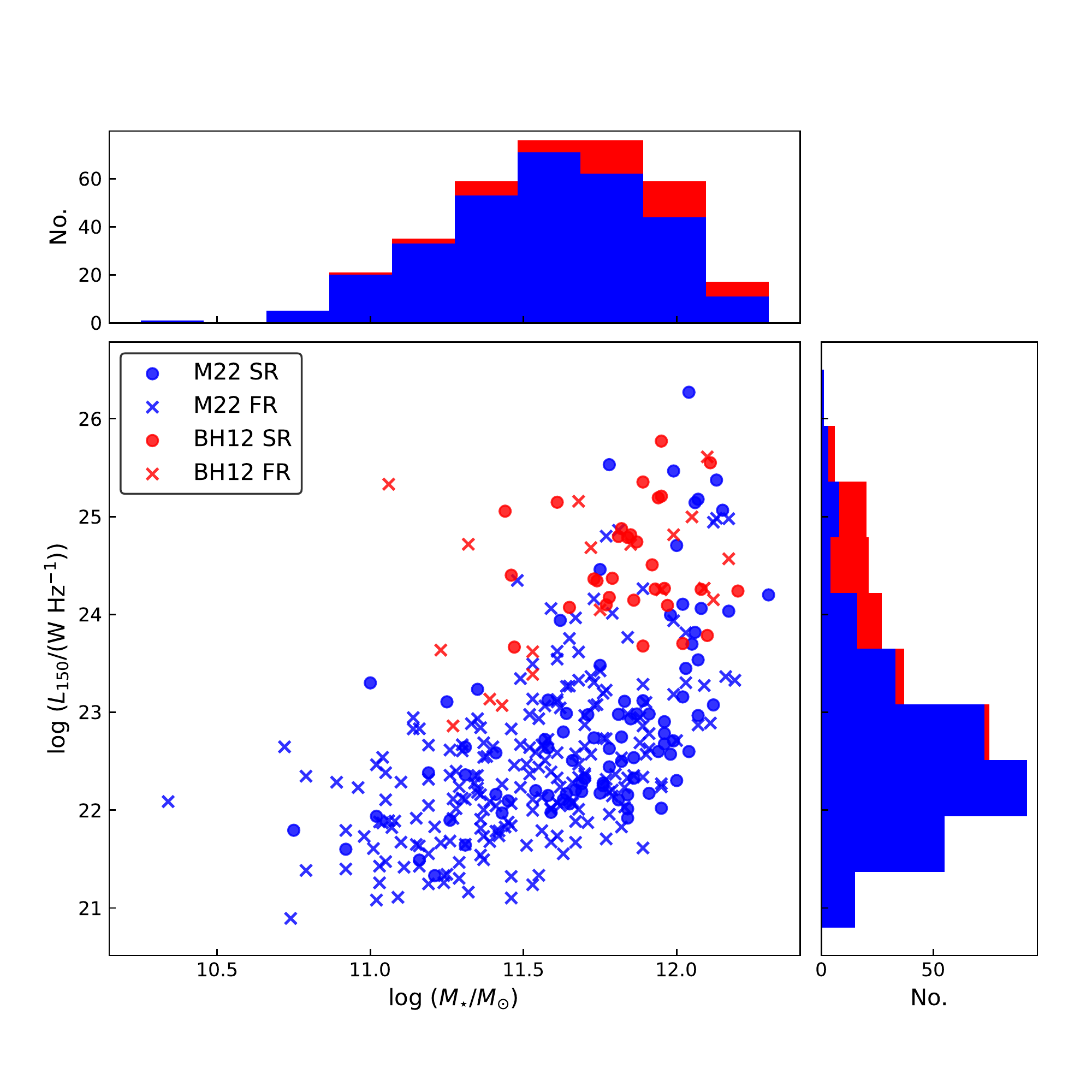}
    \caption{Radio luminosity and stellar mass distributions of the final radio AGN sample in this work.
    The blue data points and histograms represent the 300 radio AGNs from the \citetalias{Mulcahey22} sample based on LoTSS DR2.
    The red data points and bars stacked on the blue histograms denote the other 49 radio AGNs identified in \citet{Best12}.
    We use circles and crosses to mark the SRs and FRs, respectively, classified by \citet{Graham19}. 
    }
    \label{fig:lmdist}
\end{figure}

\begin{figure*}
    \centering
    \includegraphics[width=\linewidth]{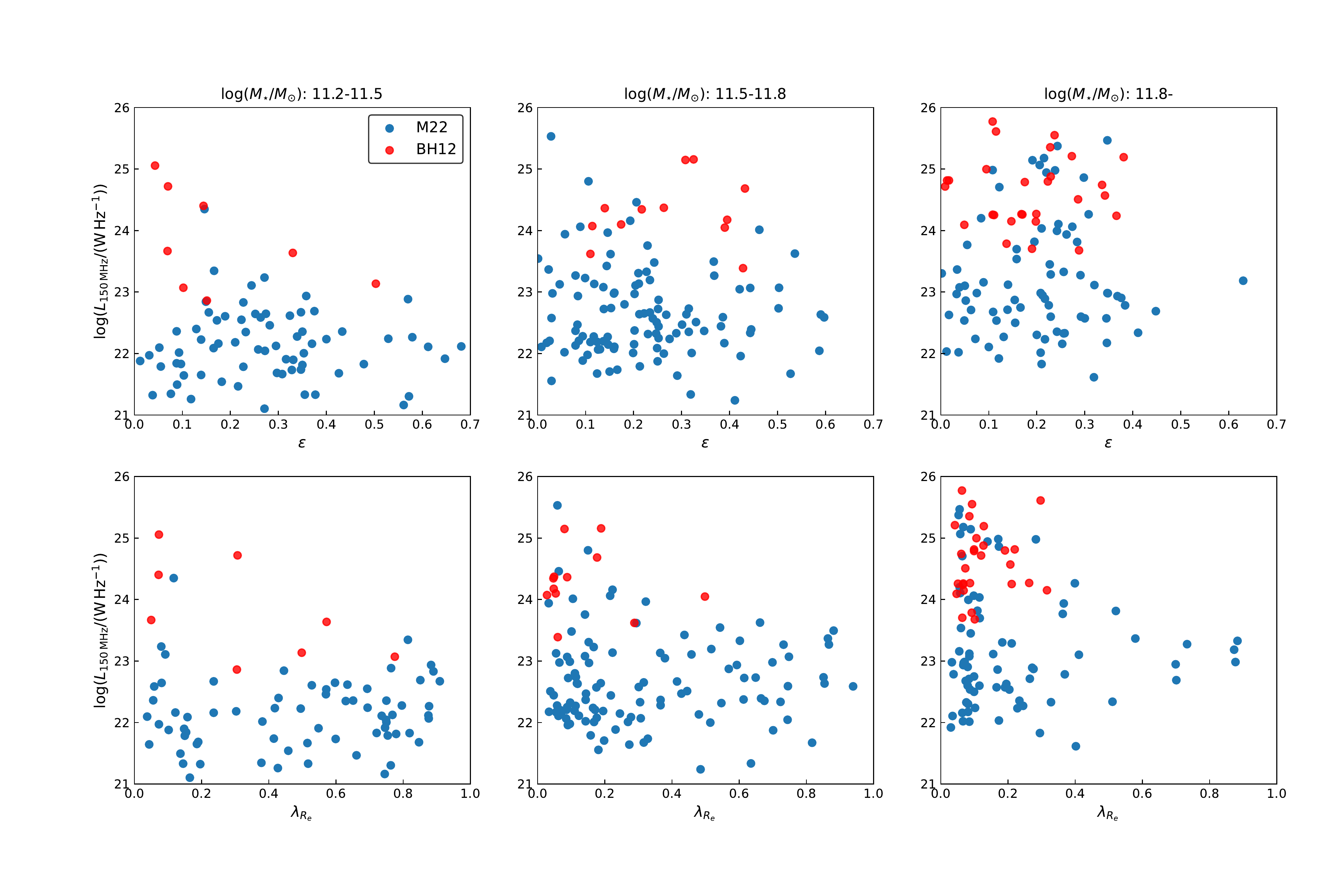}
    \caption{\radL of radio AGNs versus the $\epsilon$ and \lamre of their host galaxies. %
    The top panels show the \radL of radio AGNs as a function of the $\epsilon$ of the host galaxies in three \mstar bins (from left to right): ${\rm log} (M_{\star}/M_{\odot})$ $=11.2-11.5$, $11.5-1.8$, and $>11.8$.
    The bottom panels show the \radL as a function of the \lamre of the host galaxies in three \mstar bins.
    In all panels, the blue points denote radio AGNs from \citetalias{Mulcahey22}, and the red points denote radio AGNs from the \citetalias{Best12} sample.
    }
    \label{fig:lmor}
\end{figure*}
We used the cross-matching results from \citet[]{Mulcahey22}, hereafter `M22', to classify the radio AGNs in the MaNGA sample.
\citetalias{Mulcahey22} used a 5"-matching radius to match the positions in the source catalogues of the MaNGA and LoTSS DR2.
This matching resulted in a radio sample of 1410 MaNGA sources with a false positive rate of  less than 10\%.
At low luminosity, the contribution of star-forming galaxies in a radio source sample becomes important \citep{Best12,Sabater19}.
Therefore, M22 separated these radio sources into radio AGNs and star-forming galaxies based on four diagnostics: the $D_n4000-L_{150\,\rm MHz}/M_{*}$ diagram; the [NII]  Baldwin, Phillips, and  Telervich \citep[BPT;][]{Baldwin81} diagram; $L_{H\alpha}-L_{150\,\rm MHz}$; and the colour from the Wide-Field Infrared Survey Explorer \citep[WISE;][]{Wright10}.
This classification process was performed for LoTSS DR1, and the contamination rate from star-forming galaxies was estimated to be less than 3\% \citep{Sabater19}.
Combining the results from M22 and G19, we obtained a clean MaNGA-radio AGN sample that consists of 300 radio AGNs with reliable stellar kinematic measurements.
We also obtained a total of 2796 galaxies in G19's clean sample that fall into the field of view (FoV) of the LoTSS DR2.

We note that the simple cross-matching process may miss some luminous radio AGNs that can have a large angular size in the sky.
Moreover, some MaNGA sources are outside the FoV of LoTSS DR2.
Therefore, we also matched the MaNGA sample with the radio sources catalogue provided in \citet[]{Best12}, hereafter `BH12', which is based on the NVSS-FIRST data and the SDSS data release 7 \citep[DR7;][]{Abazajian09}.
This added 1090 galaxies covered by the FoV of SDSS DR7 from G19's clean sample.
The radio catalogue from BH12 used a more careful hybrid cross-matching method in the identification of optical counterparts \citep{Best05a,Donoso09}. 
This dataset has a radio flux limit of 5 mJy and mainly contains radio sources with $L_{1.4\,\rm GHz}>10^{23}\,\rm W\,Hz^{-1}$.
The radio AGNs in this catalogue and in M22 are classified in a similar way \citepalias[see Appendix 1 in][]{Best12}.
In this way, we obtained 49 more radio AGNs with kinematic measurements.
%
Therefore, our final sample contains 3886 galaxies, of which 349 are radio AGNs.

The radio luminosity and stellar mass distributions of our final radio AGN sample are shown in Figure  \ref{fig:lmdist}.
The stellar masses used in this work are taken from the source catalogue of \citet{Graham19}, which is based on the calibration of the dynamical mass against the K-band luminosity of \citet{Cappellari13}.
%
The radio luminosities were determined using the redshifts taken from NSA and a canonical radio spectral index of 0.7.
The majority of the radio AGNs are within $M_{\star}=10^{11}-10^{12.5}\,M_{\odot}$ and $L_{150\,\rm MHz}=10^{21}-10^{24}\,\rm W\,Hz^{-1}$.
Over 86\% of these radio AGNs are hosted by passive galaxies based on the colour-colour criteria in \citet{Chang15} ($u-r>1.6\times(r-z)+1.1$).
Nearly all the radio AGNs identified in \citetalias{Best12} are above $L_{150\,\rm MHz}=10^{23}\rm\,W\,Hz^{-1}$ and therefore  complement the LoTSS DR2 sample, which has only a very small number of sources at high luminosity, and improve the coverage of the \radL-$z$ plane for the more luminous sources.
The rotator type of each source is also marked in Figure  \ref{fig:lmdist}.
No significant trend between the rotator type and radio luminosity was found.

\begin{figure*}
    \centering
    \includegraphics[width=\textwidth]{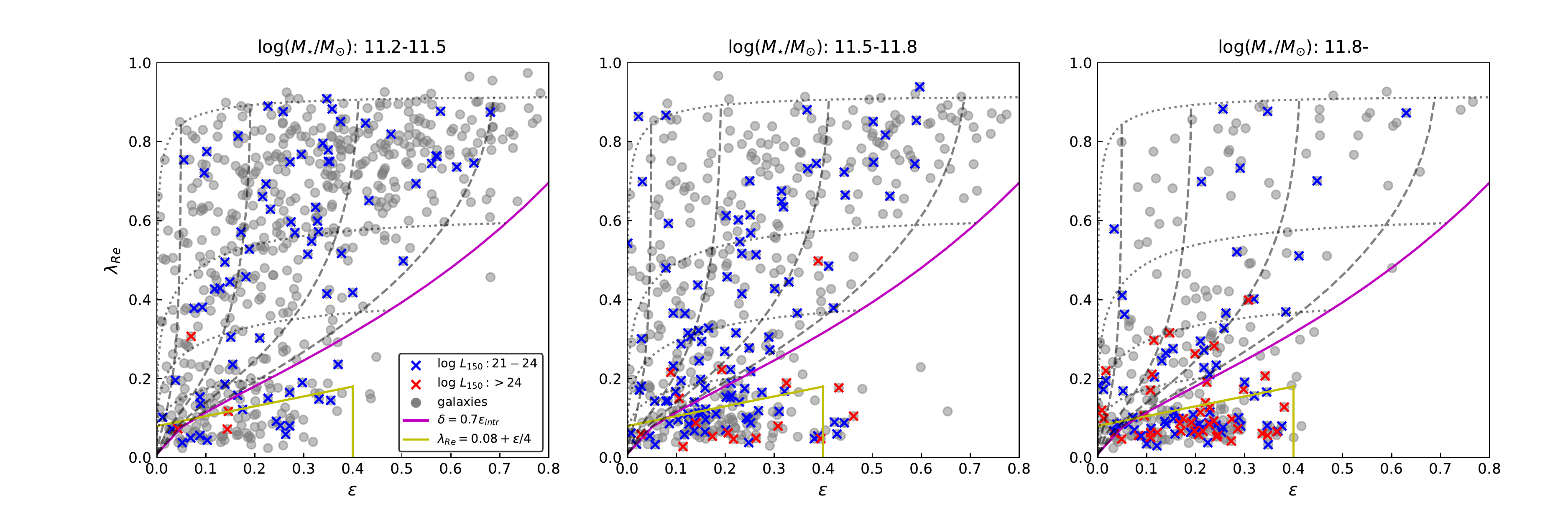}
    \caption{Distributions of galaxies and radio AGN in the $\epsilon$-\lamre\,plane.
    From left to right, the three panels show the distributions of galaxies in three \mstar bins: log($M_{\star}/M_{\odot}$)=$11.2-11.5$, $11.5-1.8,$ and $>11.8$.
    The total number of AGNs and galaxies within these \mstar ranges are 302 and 1329, respectively.
    The source number in each \mstar bin is listed in Table \ref{tab:num}.
    The blue and red crosses mark the radio AGNs with log(\radL)=$21-24$ and $>24$, and the grey points denote the galaxies within the corresponding \mstar bin. 
    The yellow lines denote the separation of SRs and FRs proposed by \citet{Cappellari16}.
    The magenta line shows the estimated \lamre of edge-on FR galaxies with maximum anisotropy $\delta=0.7\epsilon_{intr}$ described in \citet{Cappellari07}, where $\epsilon_{intri}$ denotes the intrinsic ellipticity. The grey dashed and dotted lines show the \lamre for FRs with different inclinations and $\epsilon_{intr}$, respectively.
    }
    \label{fig:lamdaep}
\end{figure*}

\begin{figure*}
    \centering
    \includegraphics[width=\textwidth]{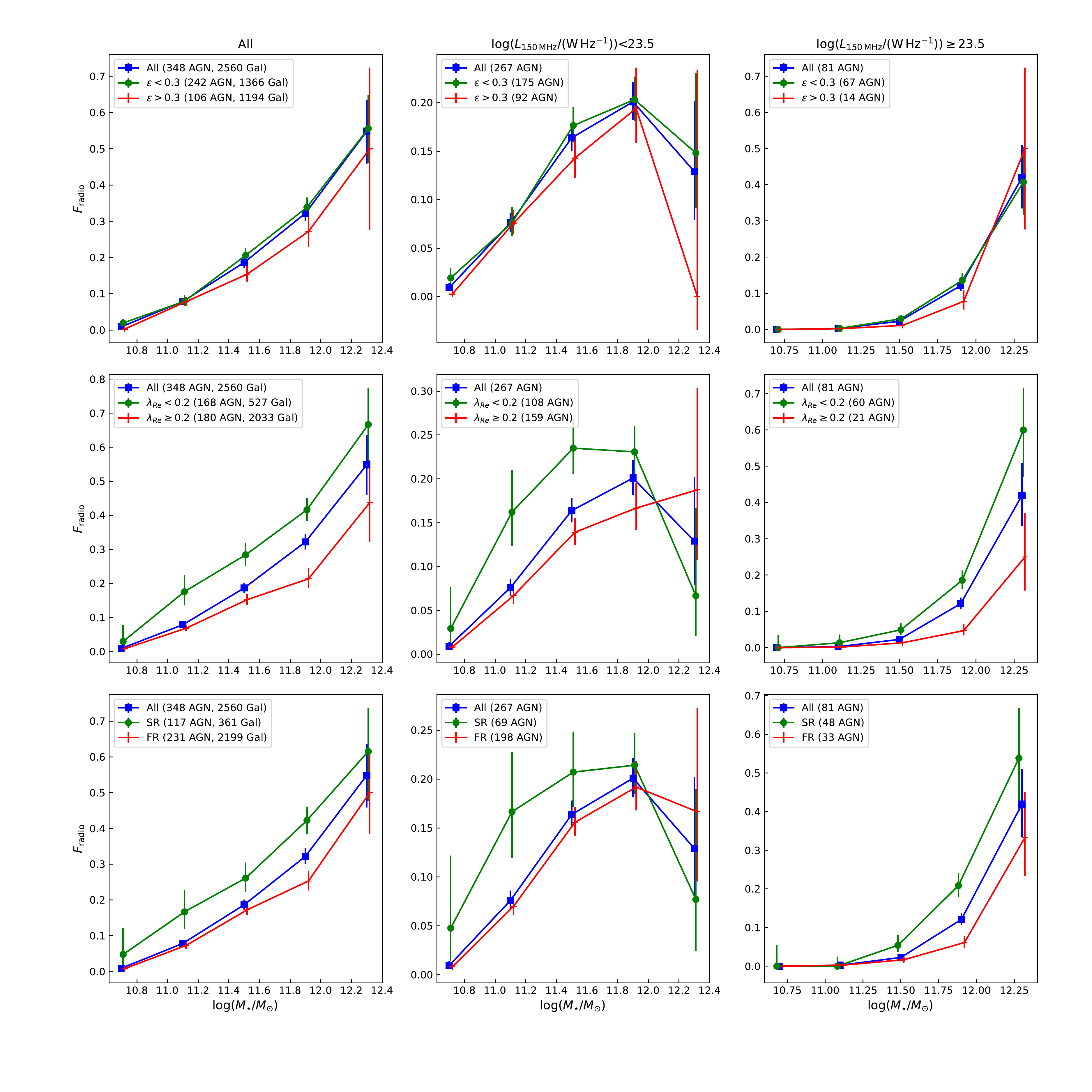}
    \caption{\frl-\mstar relations in different galaxies. %
    From left to right, the three panels in each row show the \frl for all radio AGNs, radio AGNs with \radL$<10^{23.5}\,\rm W\,Hz^{-1}$, and radio AGNs with \radL$\geq10^{23.5}\,\rm W\,Hz^{-1}$.
    Top row: \frl-\mstar relations for galaxies with different $\epsilon$. 
    The blue squares denote the \frl for all galaxies, and the green points and red dots denote the \frl for galaxies with $\epsilon<0.3$ and $\epsilon>0.3,$ respectively.
    The three lines were shifted a bit horizontally for clarity.
    Middle row: \frl-\mstar relations for galaxies with different \lamre.
    The blue squares denote the \frl for all galaxies, and the green points and red dots denote the \frl for galaxies with \lamre$<0.2$ and \lamre$>0.2$, respectively.
    Bottom row: \frl-\mstar relations for galaxies with different rotator types.
    The blue squares denote the \frl for all galaxies, and the green points and red dots denote the \frl for SRs and FRs, respectively.
    The number of AGNs used in each \frl relation is listed in the corresponding panel.
    The number of galaxies used in the \frl calculations is listed in the left panel of each row, and the \frl in the middle and right panels was calculated based on the same galaxy sample as in the left panels.
    All error bars are the 1$\sigma$ Agresti-Coull confidence limits \citep{agresti98}.
    }
    
    \label{fig:frl_m}
\end{figure*}
\section{Data analysis}\label{sec:analysis}

\begin{figure*}
    \centering
    \includegraphics[width=\linewidth]{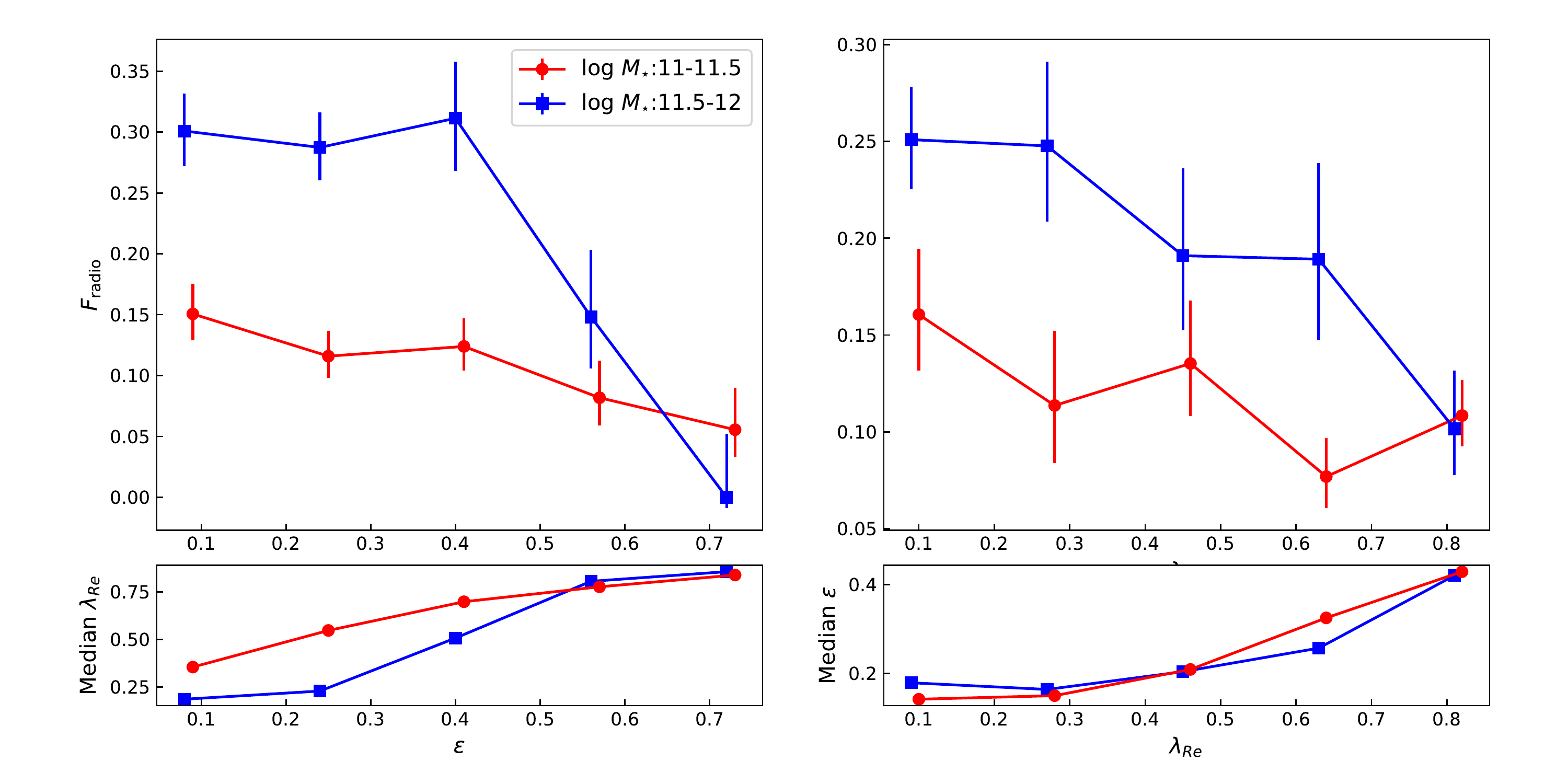}
    \caption{\frl for galaxies with different morphologies and kinematics.
    Top: \frl as a function of $\epsilon$ and \lamre for galaxies with different \mstar.
    The red dots and blue squares represent galaxies with log(\mstar$/M_{\odot}$)$=11-11.5$ and $11.5-12$, respectively.
    Bottom: Median \lamre for galaxies in each $\epsilon$ bin and the median $\epsilon$ in each \lamre bin.
    The red lines were shifted horizontally a bit for clarity.
    }
    \label{fig:frl_mor}
\end{figure*}

\subsection{The $\epsilon$ and \lamre of galaxies hosting a radio AGN} \label{sec:eplam}
Previous studies have found that the radio luminosity distribution of radio AGNs is linked with the optical projected axis ratio $q$ \citep[the ratio of the minor axis to the major axis; ][]{Zheng20,Zheng22}.
Galaxies with a large $q$ ($>$0.6) can host a high-power radio AGN with \radL$\gtrsim10^{23.5}\rm\,W\,Hz^{-1}$, while galaxies with a small $q$ host radio AGNs mostly with a lower radio luminosity.
%
This trend can be seen in the radio AGNs' distribution on the \radL-$q$ plane, where most radio AGNs are distributed in a triangle area and leave the small-$q$-high-\radL region empty \citep[see Figure 3 in ][]{Zheng20}.

\begin{table}
    \centering
    \begin{tabular}{c|ccc}
         log(\mstar/$M_{\odot}$)& 11.2--11.5 & 11.5--11.8 & >11.8 \\\hline
         LoTSS AGN&  70 & 106 & 78 \\\hline
         BH12 AGN& 8 & 12 & 28 \\\hline
         Total AGN& 78 & 118 & 106 \\\hline
         Galaxy & 575 & 468 & 286 \\\hline
    \end{tabular}
    \caption{Number of sources in the three \mstar bins in Section \ref{sec:eplam}.
    The number of AGNs from the LoTSS DR2 and BH12 are listed separately. 
    The number of galaxies is based on the G19's clean galaxy sample in the FoVs of either LoTSS DR2 or SDSS DR7.}
    \label{tab:num}
\end{table}
The radio AGNs in this work should follow the \radL-$q$ distribution found in previous works.
We selected radio AGNs from three \mstar bins with a high number of sources: ${\rm log} (M_{\star}/M_{\odot})$ $=11.2-11.5$, $11.5-11.8,$ and $>11.8$.
The total number of radio AGNs and galaxies within the \mstar ranges is 302 and 1379, respectively.
The number of sources in each of these three bins is listed in Table \ref{tab:num}.
Because \mstar is correlated to both the radio luminosity and the morphology of galaxies \citep{Chang13, Sabater19}, constraining the \mstar range is necessary to show a pure luminosity-morphology correlation.
In the catalogue from \citetalias{Graham19}, the $\epsilon$ of galaxies was provided and can be written as $\epsilon=1-q$. 
The resulting distributions of radio AGNs in the \radL-$\epsilon$ plane in the three \mstar bins are shown in the upper panels of Figure \ref{fig:lmor}.
%

In the first \mstar bin, while 13 of the 72 radio AGNs with \radL$<10^{23.5}\rm\,W\,Hz^{-1}$ have a large ellipticity $\epsilon>0.4$, none of radio AGNs with higher luminosity have a large ellipticity.
This means that a high-power radio AGN is not likely to reside in an elongated galaxy, which is in line with previous studies \citep{Barisic19,Zheng20}.
We note that this difference between high- and low-power radio AGNs does not seem to be significant in higher \mstar bins.
This is mainly because most of the massive galaxies are round elliptical galaxies with a small $\epsilon$. 
In our sample, more than 90\% of the galaxies with \mstar$>10^{11.8}\,M_{\odot}$ have an $\epsilon$ smaller than 0.4.

The apparent ellipticity of galaxies can greatly differ from the intrinsic ellipticity because of projection effects \citep{Binney85}.
Therefore, the observed $\epsilon$ is not an ideal proxy for the kinematics of galaxies.
To better describe the kinematic features of galaxies, we use \lamre, which quantifies the importance of the regular motions with respect to the random motion of stars in galaxies by definition.

We plotted the radio AGNs in the three \mstar bins on the \radL-\lamre\, plane, shown in the bottom panels of Figure \ref{fig:lmor}.
The distribution of the radio AGNs is clearly characterised by a triangle zone with nearly no radio AGNs appearing in the large-\lamre-high-\radL area.
The maximum \radL of radio AGNs decreases dramatically with \lamre\, in all three \mstar bins.
In the first \mstar bin, the fraction of large \lamre (\lamre>0.4) sources is 0/4 for radio AGNs with \radL$>10^{24}\,\rm W\,Hz^{-1}$ and 47/74 for those with lower luminosity.
These fractions become 1/15 versus 35/102 in the second bin and 0/41 versus 11/65 in the third bin. 
This result indicates that the radio AGNs with \radL$>10^{24}\,\rm W\,Hz^{-1}$ in our sample are triggered in galaxies with \lamre$\lesssim0.4$.
Furthermore, the comparison between the \radL-$\epsilon $ and \radL-\lamre\, diagrams implies that the \radL-\lamre\, relation is a powerful tool for studying the intrinsic relations between radio AGNs and the kinematics of their host galaxies.

In summary, we show the location of the radio AGNs in the $\epsilon$-\lamre\, plane in Figure \ref{fig:lamdaep}.
Apparently, luminous radio AGNs tend to be in SRs, which have both a small $\epsilon$ and \lamre.
On the contrary, less luminous radio AGNs can be found in many FRs.
The fractions of SRs in high-power AGNs are $50^{+22}_{-22}\%$,\footnote{The errors of these fractions show the 1$\sigma$ Agresti-Coull confidence limits\citep{agresti98}.} $56.3^{+11.7}_{-12.4}\%$, and $68.3^{+6.8}_{-7.6}\%$ in the three bins, while the fractions in low-power AGNs are $16.2^{+4.8}_{-3.9}\%$, $24.5^{+4.5}_{-4.0}\%$, and $53.8^{+6.1}_{-6.2}\%$.
In all three bins, high-power radio AGNs have a significantly higher fraction of SRs in contrast to low-power radio AGNs.
Because SRs have been argued to have a dry-merger dominant history, in contrast to FRs, we suggest that the difference in the $\epsilon$-\lamre\, distributions of high- and low-luminosity radio AGNs at fixed \mstar stems from the different evolutionary paths of the host galaxies.

\subsection{Prevalence of radio AGNs}
%
%
When inspecting the distributions of galaxies and radio AGNs in the $\epsilon$-\lamre\, plane in Figure \ref{fig:lamdaep}, we found that more massive galaxies have a higher fraction of radio AGNs and have a higher radio luminosity.
This is in line with previous studies based on large radio surveys \citep[e.g.][]{Best05b,Sabater19} that showed that the fraction of galaxies hosting a radio AGN, $F_{\rm radio}$, is tightly related to stellar mass and radio luminosity.
Figure\ref{fig:lamdaep} also shows that the fraction of high-luminosity radio AGNs seems to decrease with either \lamre or $\epsilon$, which is consistent with the correlations between $F_{\rm radio}$ and the shape of galaxies in \citet{Barisic19} and \citet{Zheng20}.
In this section, we investigate these trends with a focus on how the kinematics of galaxies may influence $F_{\rm radio}$.

\begin{figure}
    \centering
    \includegraphics[width=0.9\linewidth]{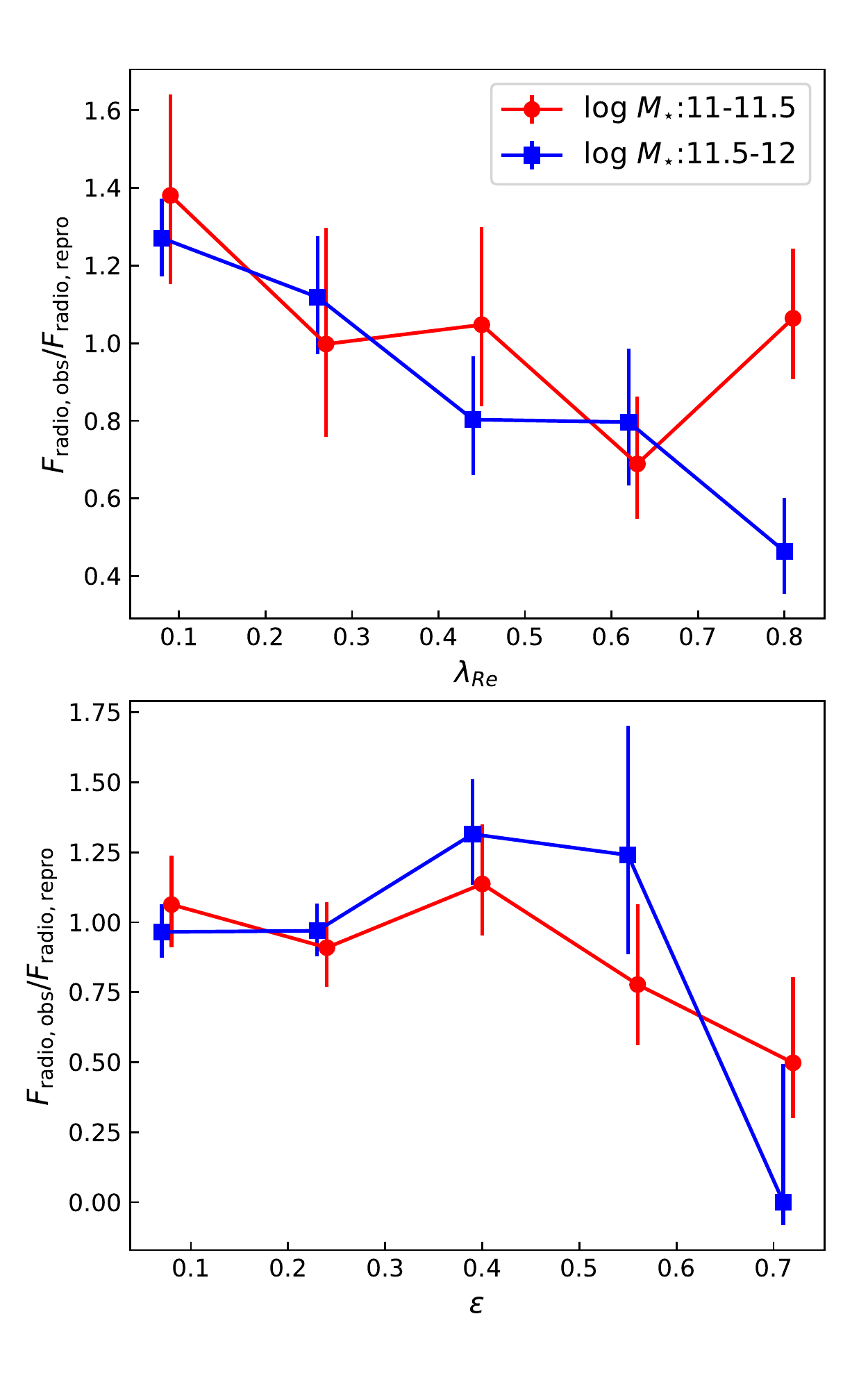}
    \caption{Ratio of the observed and reproduced \frl. {Top}: \frl ratio as a function of \lamre. The colours and symbols are similar to Figure \ref{fig:frl_mor}. {Bottom}: \frl ratio as a function of $\epsilon$. All red lines are shifted a bit horizontally. }
    \label{fig:frl_pred}
\end{figure}
We define \frl as 
\begin{equation}
   F_{\rm radio}(\theta_1,\theta_2,...)=\frac{N_{\rm radio AGN}(\theta_1,\theta_2,...)}{N_{\rm galaxy}(\theta_1,\theta_2,...)}, 
\end{equation}
where $N_{\rm radio AGN}$ is the number of radio AGNs identified in either the \citetalias{Mulcahey22} or \citetalias{Best12} sample, $N_{\rm galaxy}$ is the number of galaxies covered by the FoV of the LoTSS DR2 or the SDSS DR7, and $\theta_i$ denotes an arbitrary parameter constraint in the source selection (e.g. \mstar, \radL, \lamre).
We note that our \frl results should only be taken as an accurate measurement for the prevalence of radio AGNs if the
sensitivity variation across the FoV and the change in the detectability of certain galaxies at different redshifts are considered in the calculation.
As we do not do so, our results may differ from those of previous works \citep{Best05b,Sabater19}.
However, in this work we focus on the relative change of \frl in galaxies with different kinematics, and this change should not be influenced by these factors.
Our simple estimation of \frl should be adequate for the goals in this paper.

We first divided the sources into two groups with different $\epsilon$ and compared the \frl-\mstar\, relations within the two groups.
We choose $\epsilon=0.3$ as the division line because it is roughly where the radio luminosity distribution changes, as shown in Figure  \ref{fig:lmor}.
This $\epsilon=0.3$ division also ensures the group with a larger $\epsilon$ has enough sources for the \frl calculations.
We calculated the \frl in five logarithmic, equidistant \mstar bins between $10^{10.5}$ to $10^{12.5}\,M_{\odot}$.
The results are shown in the top-left panel of Figure  \ref{fig:frl_m}.
The error bars are the 1$\sigma$ Agresti-Coull confidence limit \citep{agresti98}.
To investigate how the radio AGN-morphology relation changes with radio luminosity, we further calculated the \frl for low- and high-luminosity radio AGNs with \radL lower and higher than $10^{23.5}\rm\,W\,Hz^{-1}$ and show the \frl  in the top-middle and top-right panels of Figure  \ref{fig:frl_m}.

The \frl for all radio AGNs in both the large- and small-$\epsilon$ groups follow  the canonical increasing trend with \mstar \citep{Best05a,Sabater19}.
The \frl of the large-$\epsilon$ galaxies are slightly smaller than those of the small-$\epsilon$ galaxies in most bins.
However, these differences do not seem to be significant, as they are typically within the 1$\sigma$ error bars.
This result is in line with the findings in \citet{Zheng20}, where the \frl of oblate galaxies are close to the \frl of triaxial galaxies.
The marginal differences between the \frl of large- and small-$\epsilon$ galaxies are also present in the results for high- and low-luminosity radio AGNs.
Although in the most massive bin the large-$\epsilon$ galaxies have a higher \frl for high-luminosity radio AGNs, the number of sources in the bin is too small for this result to be significant.

We then separated the sources into two groups based on the \lamre to investigate the \frl-\mstar relation for galaxies with different kinematics.
We chose \lamre$=0.2$ as the dividing line based on the \radL-\lamre distribution in Figure \ref{fig:lmor}.
The \frl results are shown in the middle row of Figure  \ref{fig:frl_m}.

Based on Figure  \ref{fig:frl_m}, it is apparent that the low-\lamre galaxies have a significantly higher \frl than the high-\lamre galaxies in almost all \mstar bins with reliable \frl results (small error bars).
Moreover, this trend does not change for radio AGNs with different radio luminosities.
This result shows that both low- and high-luminosity radio AGNs are more likely to be triggered in galaxies with a low-\lamre.

Finally, we also separated the sources based on their rotator type in \citetalias{Graham19}. We show the results in the bottom row of Figure \ref{fig:frl_m}.
The results are largely similar to those of the \lamre groups.
This is because the SRs are all low-\lamre galaxies by definition.

To understand in more detail the \frl dependence on galaxy morphology and kinematics, we also calculated the \frl as a function of either $\epsilon$ or \lamre.
We chose galaxies with \mstar values ranging from $10^{11}$ to $10^{12}\,M_{\odot}$ where the sample has a large source number and shows the largest difference in \frl for galaxies with different $\epsilon$ or \lamre in Figure \ref{fig:frl_m}.
To enlarge the source number in each bin and reduce the error bars of the \frl, we split the sources into two \mstar groups, one with log(\mstar$/M_{\odot}$)=$11-11.5$ and another with$11.5-12$, and we calculated the \frl in five equidistant $\epsilon$ or \lamre bins.
We only included sources with $\epsilon=0-0.8$ and \lamre$=0-0.9$ because the number of sources with large $\epsilon$ and \lamre is very low.
The number of AGNs in these two groups is 108 and 186, respectively, and the number of galaxies is 902 and 667, respectively.
The \frl results are shown in the top panels of Figure  \ref{fig:frl_mor}.

For the \frl-$\epsilon$ relation, the \frl shows a weak decreasing trend in the low-\mstar group.
In the high-\mstar groups, the \frl does not change until $\epsilon=0.4$, where it then drops dramatically.
This may explain the marginal differences in the \frl for the large- and small-$\epsilon$ galaxies in Figure \ref{fig:frl_m}, where a division line of $\epsilon=0.3$ was adopted.
We note that the \frl-$\epsilon$ relation for the high-\mstar galaxies is not directly in line with the \frl-$q$ relations in \citet{Barisic19} or \citet{Zheng20}, as the \frl-$q$ relation in \citet{Barisic19} did not have a plateau and the \frl was not dependent on $q$ in \citet{Zheng20}.
A likely reason for this difference is that the three works use different sample criteria.
Both \citet{Barisic19} and \citet{Zheng20} used mass-limited quiescent (non-star-forming) galaxy samples in their analyses, while \citet{Zheng20} used radio AGNs with a much lower luminosity limit than \citet{Barisic19}.
In our work, to enlarge the sample size, we included as many sources as possible regardless of \mstar limits, colours, or radio luminosity limits.
Therefore, the \frl-$\epsilon$ relation in this work is not, or only partially, in line with previous works.

In the top-right panel of Figure  \ref{fig:frl_mor}, the \frl of high-\mstar galaxies decreases monotonically with \lamre.
For low-\mstar galaxies, the \frl-\lamre\, relation also shows a weak, decreasing trend despite small fluctuations.
Based on these results, we conclude that radio AGNs are triggered more easily in galaxies with stronger random stellar motions (i.e. galaxies that have weaker stellar disc components).

\subsection{Observed vs. reproduced \frl relations}
In the previous section, we show that \frl is anti-correlated with both $\epsilon$ and \lamre.
However, the morphologies and kinematics of galaxies are related to each other.
For FRs, as shown in Figure \ref{fig:lamdaep}, a large $\epsilon$ typically indicates the presence of a prominent disc component and a large \lamre\, \citep[see Eqs. 14-18 in][]{Cappellari16}.
Above a characteristic mass of about $2\times10^{11}\,M_{\odot}$, galaxies start to be dominated by SRs, which have been assembled through dry mergers \citep[see review][Sec. 7]{Cappellari16}.
The kinematics of SRs tend to be dominated by random motions.
They do not have stellar discs and intrinsically have $\epsilon\lesssim0.4$ in projection. 
Therefore, when plotting the median \lamre or $\epsilon$ in each bin of Figure \ref{fig:frl_mor} as a function of the other parameter in the bottom panels of Figure \ref{fig:frl_mor}, we observed clear increasing trends.
These trends imply that the two \frl relations actually reflect the same intrinsic link between jet-triggering probability and galaxy structure or evolutionary path.
In this section, we discuss which parameter is the better proxy of the intrinsic \frl\ relation.

Because the two \frl\ relations are connected by the correlation of $\epsilon$ and \lamre, one of the \frl relations may be reproduced by the other.
We first assumed that $\epsilon$ is a better predictor of \frl, and we expected that the \frl-\lamre\, relations could be reproduced by the \frl-$\epsilon$ relation.
We interpolated the observed \frl-$\epsilon$ relation to obtain a continuous function $\hat{F}_{\rm radio}(\epsilon)$.
This function served as input as the `intrinsic' probability of a galaxy with $\epsilon$ hosting a radio AGN.
Next, we obtained the expected radio-loud probabilities for galaxies in each \lamre bin based on their $\epsilon$ and $\hat{F}_{\rm radio}(\epsilon)$.
The average of these expected probabilities was taken as the $\epsilon$-reproduced fraction of radio AGNs $F_{\rm radio,repro}$ in the corresponding \lamre\, bin.
We changed the role of $\epsilon$ and \lamre\, in order to obtain the \lamre-reproduced \frl in each $\epsilon$ bin.
We present the ratio between the observed fraction of radio AGNs $F_{\rm radio,obs}$ and $F_{\rm radio,repro}$ in Figure \ref{fig:frl_pred}.

If any of the \frl relations can be predicted by the other \frl relation and the $\epsilon$-\lamre\, relation, we expected that the reproduced \frl would be close to the observed value, and the decreasing trend in the \frl relation would disappear in the \frl ratio diagrams.
In the top panel of Figure \ref{fig:frl_pred}, the \frl ratio $F_{\rm radio,obs}/F_{\rm radio,repro}$ for the high-\mstar sample still shows a decreasing trend with \lamre.
%
This means that the \frl-$\epsilon$ relation cannot reproduce the \frl-\lamre\, relation. Therefore, $\epsilon$ is not sufficient for showing the intrinsic link between \frl and galaxy structure.
In contrast, the $F_{\rm radio,obs}/F_{\rm radio,repro}$-\lamre\, relation for the low-\mstar sample (top panel of Figure \ref{fig:frl_pred}) and the $F_{\rm radio,obs}/F_{\rm radio,repro}$-$\epsilon$ relation for both samples (bottom panel of Figure \ref{fig:frl_pred}) do not simply decline.
This implies that \lamre might be a good proxy for the intrinsic link between \frl and galaxy structure in both samples.
In the low-\mstar sample, $\epsilon$ might also be used as a proxy, but it would be less robust than \lamre.  

We note that the binning of data may weaken the significance of the analysis, and therefore we also compared the observed and reproduced cumulative distribution of radio AGNs.
The observed cumulative distribution of radio AGNs as a function of parameter $\theta$ ($\epsilon$ or \lamre) $Cum(\theta)$ is the number of radio AGNs with $\theta_i$ less than a given value as below:
\begin{equation}
    Cum(\theta) = \sum_{i}^{N_{\rm AGN}} H(\theta-\theta_i),
\end{equation}
where $H(x)$ is the Heaviside step function and the summation is for all radio AGNs.
The reproduced cumulative distribution of radio AGNs as a function of \lamre\, using $\hat{F}_{\rm radio}(\epsilon)$ can be calculated as
\begin{equation}
    Cum_{\rm repro}(\lambda_{R_e}) = \sum_{i}^{N_{\rm gal}} \hat{F}_{\rm radio}(\epsilon_i) H(\lambda_{R_e}-\lambda_{R_e,i}),
\end{equation}
where the summation is over all galaxies in the sample.
Similarly, $Cum_{\rm repro}(\epsilon)$ can be obtained using $\hat{F}_{\rm radio}(\lambda_{R_e})$.

\begin{figure*}
    \centering
    \includegraphics[width=\textwidth]{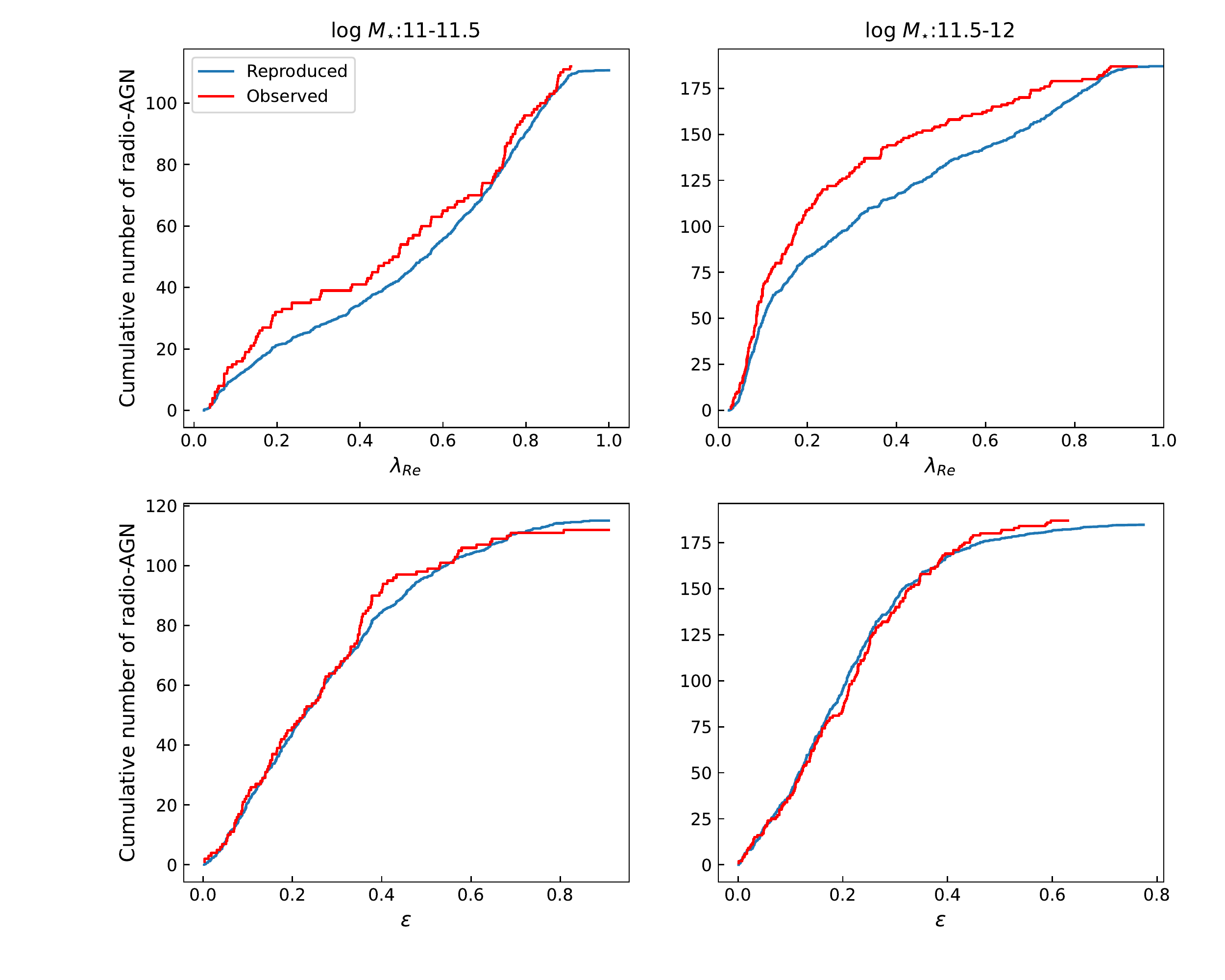}
    \caption{Cumulative distributions of radio AGNs.
    {Top:} Comparisons between the observed cumulative distributions of radio AGNs as a function of \lamre and the $\epsilon$-reproduced distributions.
    {Bottom:} Comparisons between the observed cumulative distributions of radio AGNs as a function of $\epsilon$ and the \lamre-reproduced distributions.
    }
    \label{fig:cum}
\end{figure*}

The cumulative distributions of radio AGNs for the two \mstar groups are shown in Figure \ref{fig:cum}.
The reproductions of the cumulative distributions clearly have different levels of goodness.
%
For $Cum(\lambda_{R_e})$, the distributions reproduced from $\hat{F}_{\rm radio}(\epsilon)$ are significantly different from the observed distributions.
On the contrary, the reproduced cumulative distributions $Cum_{\rm repro}(\epsilon)$ based on $\hat{F}_{\rm radio}(\lambda_{R_e})$ fit the observed distributions well. 

These results show that the \frl-\lamre\, relation describes the intrinsic links between radio jets and host galaxies better than the \frl-$\epsilon$ relation.
Because \lamre\, is a better proxy of the stellar kinematics of galaxies, this implies the triggering of radio AGNs is related to galaxy kinematics and assembly history.
Furthermore, the \frl-\lamre\, and \frl-$\epsilon$ relations should reveal the same underlying relation between radio AGNs and host galaxies because the \frl-$\epsilon$ relation can be reproduced by the \frl-\lamre\, relations.
However, $\epsilon$ is more easily affected by the inclination of galaxies, and 
this should explain the inability of the \frl-$\epsilon$  to predict the \frl-\lamre\, well.
%

%

\section{Conclusion and discussion}\label{sec:con}
In this work, we combined the morphology and kinematic measurements for galaxies in the MaNGA survey with the data of radio AGNs from LoTSS DR2 and the NVSS-FIRST survey.
We investigated the relations between radio AGNs and the kinematics of their host galaxies in \mstar constrained samples.
The main new findings in this article are briefly listed below:
\begin{itemize}
    \item The radio luminosity of a radio AGN depends on the \lamre of the host galaxy in our sample. Higher-luminosity radio AGNs could be hosted by galaxies with lower \lamre at fixed \mstar.
    \item The fraction of galaxies hosting a radio AGN is higher in galaxies with smaller \lamre in our sample. This trend is similar for both high- and low-\radL radio AGNs.
    \item The \frl-\lamre\, relation cannot be well reproduced using the \frl-$\epsilon$, while the \frl-$\epsilon$ relation can be reproduced by the \frl-\lamre\, relation. This shows that the \frl-\lamre\, relation better describes the links between radio AGNs and host galaxies in our sample.
\end{itemize}
The dependence on \lamre\, for radio AGNs in our work implies that the build-up history of galaxies could be related to the presence of radio jets.
Because \lamre\, quantifies the importance of the regular circular motion in a galaxy (in contrast to  random motion), it is an indicator of the evolutionary path of the galaxy \citep{Emsellem07,Emsellem11,Cappellari16}.
For galaxies built-up mainly from secular star forming processes, the ordered stellar motions are retained, and the stellar disc components are dominant in the galaxy structure; thus, the resulting \lamre\, would not be very small.
These galaxies are FRs. 
In contrast, SRs are likely to have a dry-major-merger dominated history, which would have led to the destruction of stellar discs and the enhancement of the random motions (i.e. a small \lamre).
Therefore, the fact that high-\radL radio AGNs favour small-\lamre\, galaxies suggests that a merger history enhances the radio AGN activities.

We stress that this does not mean that all current high-luminosity radio AGNs are triggered directly by ongoing merger events, but a merger event in the past would make a galaxy become a breeding ground for powerful radio jets. 
The heritage of a merger event may influence a radio AGN's activity in at least two ways.
Studies based on deep imaging \citep{Tadhunter16,Pierce19,Ramos11} have shown that most of the galaxies with a radio-powerful AGN have past interaction signatures.
The existence of these signatures may enhance gas inflow in the galaxy, facilitating fuelling of the central SMBH.
However, this may not be the only reason for the \radL-\lamre\, dependence because secular processes may also form structures that benefit the gas inflow and could be responsible for all radiative mode AGNs \citep[see e.g.][for more details]{Heckman14,Sellwood14}.
Another consequence of a major merger event is the impact on the black hole spin.
As mentioned in Section \ref{sec:intro}, it was suggested that a major merger could be an efficient way to make a high-spin black hole \citep{Wilson95,Moderski96,Sikora07,Fanidakis11,Barisic19} because the orbital angular momentum of the two progenitor SMBHs would transfer to the spin of the merged black hole.
However, this theory might not be fully correct.
Numerical simulations have shown that SMBHs reach their highest spin through coherent accretion processes at high redshifts when they have enough gas supply, and the merging of SMBHs at low redshifts would decrease the spin \citep{Dubois14,Bustamante19}.
The observational data also indicate that massive SMBHs might rotate more slowly than low-mass SMBHs\citep{Reynolds13}.
These results would suggest that mergers do not enhance jet-launching processes by making a high-spin SMBH.
It is more likely that mergers change the direction of the spin of SMBHs and influence the probability of launching a radio jet.
\citet{Garofalo10,Garofalo20} indicated that radio jets are more powerful in counter-rotating disc systems, which can only exist in post-merger black holes.
A counter-rotating disc system can also help form a magnetically arrested disc, which is crucial for launching a powerful jet\citep{Tchekhovskoy12,Sikora13,Rusinek20}.
Therefore, we suggest that mergers can lead to a change in the black hole spin direction and boost the probability of producing a powerful jet. 
%
%
As a result, small-\lamre\, galaxies (or SRs) can launch a powerful jet more easily than large-\lamre\, galaxies (or FRs).


%
The \frl-$\epsilon$ relation shows an overall decreasing trend but with a plateau at the small-$\epsilon$ part. 
We note that this trend is not fully consistent with previous research on the axis ratio of galaxies \citep{Barisic19,Zheng20}. 
A possible reason for this disparity is the difference in selection criteria of these works.
%
Previous works were limited to colour-selected quiescent galaxies, which would exclude a large fraction of discy galaxies.
This limitation causes large uncertainties in the \frl results for the large-$\epsilon$ (small-$q$) galaxies.
%
%
However, the sample studied in this work is mainly composed of passive galaxies, and the \frl-$\epsilon$ relation in this work does not change significantly when only the passive galaxies are included.
%
%
The \mstar limit adopted in the sample selection of \cite{Barisic19} and \cite{Zheng20} may also be a reason for our different results.

The link between the evolutionary path of galaxies and the presence of radio jets might also be related to the environment.
First, environment richness is correlated with many important properties of galaxies.
Galaxies in clusters tend to be more massive and are more likely to be elliptical galaxies than those in the field \citep[see ][and references therein]{Blanton09,Conselice14}.
Slow rotators also favour high density environments \citep{Cappellari11b}.
Second, the environment is connected to the prevalence and power of radio jets.
Galaxies in cluster environments have a higher probability to be radio-loud than galaxies with a similar stellar mass in the field \citep{Best05b,Best07,Sabater13}.
Radio AGNs also tend to have higher luminosity in higher density environments \citep{Donoso09,Croston19}.
These results and the correlations between radio AGNs and galaxy kinematics in our work can be understood within the same scenario.
We suggest the environmental trend of radio AGNs might be a result of the correlation between radio AGNs and galaxy kinematics.
Galaxies in higher density environments are more likely to have a merger-dominant evolutionary history; thus, they are also more likely to be SRs, or low-\lamre\, galaxies \citep{Cappellari11b,Cappellari16}.
Based on the results in our work, it is therefore not surprising that we can find more high-power radio AGNs in higher density environments.
%

Our work shows the potential of IFS surveys for studying the radio AGN-triggering mechanism.
To better understand the link between radio AGNs and galaxy kinematics, it will be important to increase the sample in the next generation of IFS surveys.
It will also be helpful to study the gas kinematics in these radio galaxies with deeper observations, which may provide important insights on the fuelling of the SMBHs. 
%
%

\begin{acknowledgements}
We thank Celia Mulcahey and Sarah Leslie for providing the MaNGA-LoTSS AGN sample. LOFAR data products were provided by the LOFAR Surveys Key Science project (LSKSP; https://lofar-surveys.org/) and were derived from observations with the International LOFAR Telescope (ILT). LOFAR \citep{vHaarlem13} is the Low Frequency Array designed and constructed by ASTRON. It has observing, data processing, and data storage facilities in several countries, that are owned by various parties (each with their own funding sources), and that are collectively operated by the ILT foundation under a joint scientific policy. The efforts of the LSKSP have benefited from funding from the European Research Council, NOVA, NWO, CNRS-INSU, the SURF Co-operative, the UK Science and Technology Funding Council and the Jülich Supercomputing Centre. XCZ acknowledges support from the CSC (China Scholarship Council)-Leiden University joint scholarship program.
\end{acknowledgements}

%
%
\bibliographystyle{aa}
\bibliography{LOFARMANGA}





   
  



\end{document}